\documentclass[10pt,nofootinbib,
notitlepage,superscriptaddress,twocolumn]{revtex4-1}
\pdfoutput=1

\usepackage[utf8]{inputenc} 		
\usepackage[ngerman,english]{babel} 
\usepackage[T1]{fontenc}    		
\usepackage{outlines}
\usepackage{amsmath}
\usepackage{graphicx}
\usepackage{xcolor}
\usepackage[caption=false]{subfig}
\usepackage{amssymb} 
\usepackage[pdftex,bookmarks,plainpages=false,pdfpagelabels]{hyperref}

\newcommand{\Mpl}{M_\text{Pl}}
\newcommand{\eV}{\,\mathrm{eV}}

\begin{document}

\title{Gravitational Wave Oscillations in Bigravity}

\author{Kevin Max}
\email{kevin.max@sns.it}
\affiliation{ Scuola Normale Superiore and INFN Pisa,\\ 
Piazza dei Cavalieri, 7 - 56126 Pisa, Italy}

\author{Moritz Platscher}
\email{moritz.platscher@mpi-hd.mpg.de}
\affiliation{Max-Planck-Institut f\"ur Kernphysik, \\
Saupfercheckweg 1, 69117 Heidelberg, Germany}

\author{Juri Smirnov}
\email{juri.smirnov@fi.infn.it}
\affiliation{INFN divisione di Firenze, Dipartimento di Fisica, Università di Firenze,\\
Via Sansone 1, 50019 Sesto Fiorentino, Florence, Italy}

\begin{abstract}
\noindent
\begin{center}
\textbf{ABSTRACT}
\end{center}

\noindent We derive consistent equations for gravitational wave oscillations in bigravity. In this framework a second dynamical tensor field is introduced in addition to General Relativity (GR) and coupled such that one massless and one massive linear combination arise. Only one of the two tensors is the physical metric coupling to matter, and thus the basis in which gravitational waves propagate is different from the basis where the wave is produced and detected. Therefore, one should expect -- in analogy to neutrino oscillations -- to observe an oscillatory behavior. We show for the first time how this behavior arises explicitly, discuss phenomenological implications and present new limits on the graviton parameter space in bigravity.
\end{abstract}

\maketitle

\section{Introduction}

The question whether a theory of a massless spin two particle can have a consistent massive extension was a longstanding open problem. The quest that lead to formulate this theory took place in the second half of the last century~\cite{Fierz:1939ix,Fierz:1939zz,vanDam:1970vg,Zakharov:1970cc,Vainshtein:1972sx,Boulware:1973my}. Only recently it has been proven that a consistent framework of massive gravity exists and relies on the existence of multiple spin two fields with non-linear interactions~\cite{deRham:2010ik,deRham:2010kj,deRham:2011rn,Hassan:2011hr,Hassan:2011vm,Hassan:2011tf,Comelli:2012vz,Deffayet:2012nr,Deffayet:2012zc,deRham:2016plk}.

In this letter, we study a setup with two dynamical spin two fields corresponding to two metrics~\cite{Hassan:2011zd,Hassan:2011ea} known as \textit{bigravity}. The coupling of the metrics to matter is a delicate problem and has been discussed in~\cite{deRham:2014fha} as an arbitrary choice of coupling reintroduces inconsistencies. Demanding the absence of a ghost in the theory translates into an asymmetric coupling of the metrics to matter, and this asymmetry is at the core of the physical phenomenon we will discuss in this letter. The simplest choice of matter coupling which permits a ghost-free theory is minimal coupling of one metric tensor to matter, which we will call the physical metric and no coupling of the second metric tensor to matter. This second metric tensor is a reference or sterile metric which only interacts with the physical metric via the non-linear terms in the Lagrangian. This situation is analogous to the introduction of a sterile neutrino which carries no electroweak charges.

In this theory the gravitational interactions are mediated by two gravitons, one massless and one massive. Since the two are superpositions of the physical and the sterile metric, their effective coupling to matter is different and depends on the mixing angle between the metrics. This leads to an oscillation phenomenon, first mentioned in~\cite{Berezhiani:2007zf} in a theory of massive gravity and~\cite{Hassan:2012wr} in bigravity. In this work we will study the propagation of gravitational waves (GW) in this bi-metric theory which are produced in the `flavor basis' at the source, namely only as perturbations of the physical metric. Describing the wave propagation we find a close analogy to neutrino oscillation described in the wave-packet formalism.

This phenomenon is presented for the first time in a consistent approach. Attempts have been previously made in~\cite{DeFelice2014,Narikawa:2014fua}, however only in a specific setting, and leading to an unphysical result; in particular, during propagation the mode coupling to matter exhibits an enhancement of the strain in violation of (local) energy conservation. We show that in the parameter space we consider physical no such behavior is found, as one should expect in a healthy theory. The novelty of our work in the bigravity setup is that we consider graviton masses corresponding to length scales which can be probed by astrophysical tests, while the majority of prior works have focused on much smaller graviton masses, i.e.~of the order of the Hubble scale today. This approach makes it possible to confront direct detection data of GW signals as seen by the LIGO experiment~\cite{Abbott:2016blz} with the oscillation hypothesis. The corresponding parameter space of $m_g = 10^{-22}-10^{-20}$~eV and comparably large mixing angle $\theta$ is studied, in close resemblance to the effects of pure massive gravity. Note that previous studies have found instabilities which plague the parameter regime in which the graviton mass is of the Hubble scale today, and specific parameter choices are needed to obtain viable solutions \cite{DeFelice:2014nja, Akrami:2015qga}. However, for the larger graviton masses probed here, this problem is considered to be less restricitve \cite{Babichev:2016bxi}.


\section{Gravitational wave oscillations} \label{sec:GWoscillations}

In this model, the oscillation of metric perturbations is driven by classical dynamics of the Friedmann equations~\cite{DeFelice2014}. They are extracted from the Einstein field equations of bigavity~\cite{vonStrauss:2011mq},
\begin{subequations}\label{eq:Einstein}
\begin{align}
&R_{\mu\nu} -\frac{1}{2} g_{\mu\nu} R + B_{\mu\nu}(g) = \frac{1}{M_g^2} T_{\mu\nu}\,,\\
&\tilde{R}_{\mu\nu} -\frac{1}{2} \tilde{g}_{\mu\nu} \tilde{R} + \tilde{B}_{\mu\nu}(\tilde{g}) = 0\,,
\end{align}
\end{subequations}
with
$B_{\mu\nu}(g)=m^2 \cos^2(\theta) \sum_{n=0}^3 \beta_n V^{(n)}_{\mu\nu}$ and 
$\tilde{B}_{\mu\nu}(\tilde{g}) = m^2 \sin^2(\theta) \sum_{n=1}^4 \sqrt{g^{-1} \tilde{g}} \, \beta_n \tilde{V}^{(n)}_{\mu\nu}$,
$\cos^2(\theta)=M_{\text{eff}}^2/M_g^2 $,~and $\sin^2(\theta)=M_{\text{eff}}^2/M_{\tilde{g}}^2$. The $V^{(n)}_{\mu\nu},\ \tilde{V}^{(n)}_{\mu\nu}$ encode the variation of the interaction terms in the action w.r.t.~$g,\tilde{g}$. Furthermore, by applying the covariant derivatives to Eqs.~\eqref{eq:Einstein}, we obtain the conservation laws,
\begin{equation}\label{eq:Bianchi}
\nabla_\mu B^\mu_{\hphantom\mu \nu} =0, \qquad \tilde{\nabla}_\mu \tilde{B}^\mu_{\hphantom\mu \nu} =0, \qquad \nabla_\mu T^\mu_{\hphantom\mu \nu} =0,
\end{equation}
the first two of which are known as Bianchi constraints.

\subsection{Background cosmology} \label{sec:background}

We now calculate the cosmological implications on a static background. For both $g$~and~$\tilde{g}$, we assume an FRW background metric,
$\mathrm{d}s^2 = \, a(\eta)^2(-\mathrm{d}\eta^2 + \mathrm{d}\vec{x}^2)$ and 
$ \mathrm{d}\tilde{s}^2 = \, b(\eta)^2(-\tilde{c}(\eta)^2\,\mathrm{d}\eta^2 +\nolinebreak\mathrm{d}\vec{x}^2)\,$.
The lapse function $\tilde{c}(\eta)$ determines the light cone for the second metric and plays a role for the propagation speed of the massive gravitational wave excitations. This is the most general ansatz compatible with a homogeneous and isotropic Universe~\cite{Comelli:2011zm}.

Plugging this ansatz into Eqs.~\eqref{eq:Einstein} and omitting explicit dependencies yields the cosmic evolution equations,
\begin{subequations}\allowdisplaybreaks
\begin{gather}
	\frac{3}{a^2}\left(H^2+k \right) = \Lambda(y) + \frac{\rho(\eta)}{M_g^2}, \label{eq:Einstein_g}\\
	\frac{3}{b^2}\left(J^2/\tilde{c}^2+k \right) = \frac{\tilde{\rho}(y)}{M_{\tilde{g}}^2}, \label{eq:Einstein_f}
\end{gather}
\end{subequations}
where $\Lambda(y) \equiv m^2 \sin^2\theta \left[ \beta_0 + 3\beta_1 y + 3\beta_2 y^2 + \beta_3 y^3 \right]$ and $\tilde{\rho}(y) \equiv M_{\tilde{g}}^2 m^2 \cos^2\theta \left[ \beta_1 y^{-3} + 3\beta_2 y^{-2} + 3\beta_3 y^{-1} + \beta_4 \right]$. Here, a prime denotes a derivative w.r.t.~$\eta$, $y=b/a$, and $H = a'/a$ as well as $J = b'/b$ are the Hubble parameters for both metrics in conformal time.

Moreover, Eqs.~\eqref{eq:Bianchi} imply that $\rho'(\eta) = -3 H (1+\omega) \rho(\eta)$ and
	$(\tilde{c} H - J)\left[ \beta_1 y + 2\beta_2 y^2 + \beta_3 y^3 \right]\equiv (\tilde{c} H - J) \Gamma(y) = 0,$
for a perfect fluid with equation of state $P = \omega \rho$. It was shown that only the vanishing of the round brackets yields a physically meaningful solution~\cite{vonStrauss:2011mq}.
Thus, we find $J(\eta) = \tilde{c}(\eta) H(\eta)$.

Using this result, we can derive an algebraic equation for $y$, by subtracting Eq.~\eqref{eq:Einstein_g} from Eq.~\eqref{eq:Einstein_f},
\begin{equation}\label{eq:CosmoMasterEquation}
\begin{gathered}
	\beta_1 \cos^2\theta y^{-1} +(3\beta_2 \cos^2\theta - \beta_0 \sin^2\theta) +\\ 
    + (3\beta_3 \cos^2\theta - 3\beta_1 \sin^2\theta) y + \\
    + (\beta_4 \cos^2\theta - 3\beta_2 \sin^2\theta) y^2 - \beta_3 \sin^2\theta y^3 = \frac{\rho(\eta)}{M_g^2 m^2}.
\end{gathered}
\end{equation}
By assumption, $\rho$ is the density of a perfect fluid with $\omega \ge -1$, which behaves as~\cite{vonStrauss:2011mq}
\begin{equation}
	\rho(\eta) = \rho_0 
    \begin{cases}
		1 & \text{if } \omega = -1,\\
        \left(\frac{a(\eta)}{a(\eta_0)}\right)^{-3(1+\omega)} & \text{if } \omega >-1,
	\end{cases}
\end{equation}
such that any fluid of type $\omega>-1$ is diluted, i.e.\ $\rho \nobreak\to\nobreak 0$ for $\eta \to\nobreak \infty$. It is in fact sufficient to consider such densities, since any cosmological constant (CC) type of energy density may be included in the interaction terms of the bigravity theory. 

In this limit, we denote the solution of Eq.~\eqref{eq:CosmoMasterEquation} as $y_*$. An exact expression is in principle feasible, however not very enlightening. Therefore, and since we are interested  in late times, we leave $y_*$ undetermined and linearize Eq.~\eqref{eq:CosmoMasterEquation} as $y = y_* + \delta y$ to obtain,
\begin{equation}
	\delta y(\eta) = - \frac{\rho(\eta)}{3m^2 M_g^2} \frac{y_*^3}{\Gamma_*(\cos^2\theta + y_*^2\sin^2\theta)- 2 \frac{\tilde{\rho}_*y_*^4}{3m^2M_{\tilde{g}}^2}},
\end{equation}
with the short-hand notation $\Gamma_* \,{=}\, \Gamma(y_*)$ and $\tilde{\rho}_* \,{=}\, \tilde{\rho}(y_*)$.

This manipulation allows us to rewrite Eq.~\eqref{eq:Einstein_g} as 
	$a(\eta)^{-2}( H(\eta)^2 + k ) = \frac{1}{3}\Lambda_* + \frac{\rho(\eta)}{3 \Mpl^2}$
with the physical CC $\Lambda_* = \Lambda (y_*)$
and Planck mass,
$	\Mpl^2 =  \left( M_g^2 \cos^2\theta + y_*^2\sin^2\theta - \frac{2\tilde{\rho}_* y_*^4}{3m^2M_{\tilde{g}}^2\Gamma_*}\right) \left( 
	 \cos^2\theta - \frac{2\tilde{\rho}_* y_*^4}{3m^2M_{\tilde{g}}^2\Gamma_*} \right)^{-1}, $
which approaches $ M_g^2 \left(1+y_*^2 \tan^2\theta\right)$, as $\tilde{\rho}_* \to 0$, in agreement with~\cite{DeFelice2014,Narikawa:2014fua}.

Finally, we may use that $y' = \left(\frac{b}{a}\right)' = y (J - H)$ and $J = \tilde{c} \, H$ to find that,
\begin{align}\label{eq:ctilde}
	\tilde{c}(\eta) =& 1 + \frac{y'}{y H} \simeq 1 + \frac{\delta y'}{y_* H} \nonumber\\
    		\simeq& 1 - (1+\omega) \frac{\rho(\eta)}{m^2 \Gamma_* \Mpl^2} 					\frac{y_*^2}{\frac{2\tilde{\rho}_* y_*^4}					{3m^2M_{\tilde{g}}^2\Gamma_*}-\cos^2\theta}.
\end{align}
Note that $\tilde{c}$ can be both larger or less than 1, depending on the choice of the $\beta$-parameters. However, $\tilde{c} > 1$ would introduce GWs propagating with a speed larger than the speed of light. In certain frameworks this might be acceptable, e.g.~the present case is similar to the framework studied in~\cite{Babichev:2007dw}, where all matter propagates on the $g$ background and no causal paradoxes arise.

From Eq.~\eqref{eq:ctilde} we obtain $|\tilde{c} -1| \approx 10^{-20}$ for 
typical values in the parameter region of interest.
This motivates the limit where $\tilde{c} = 1$ and $y$ takes the constant value $y_*$, which we apply in the following. 



\subsection{Gravitational wave oscillations} \label{sec:oscillations}

We now address the propagation of tensor perturbations around the background metric.\footnote{The scalar mode couples to the trace of a conserved source and will thus in principle be excited, too. However, it is suppressed due to the Vainshtein effect~\cite{deRham:2014zqa}.} Defining the transverse traceless components,
%
the equations of motion are~\cite{Comelli:2012db},
\begin{subequations}
\begin{align}
h''+2 \, H h' + k^2 h + \sin^2\theta \, m^2\, \Gamma_* a^2 (h - \tilde{h}) &= 0\,,\\
\tilde{h}''+2 \,H \tilde{h}' + k^2 \tilde{h} + \cos^2\theta \, \frac{m^2 \, \Gamma_*}{y_*^2} a^2 (\tilde{h} - h) &= 0\,,
\end{align}
\end{subequations}
where $k=|\vec{k}|$ denotes the three-momentum and the polarization indices $+/\times$ are implicit. For the linear combinations
$h_1 \equiv \cos^2\theta \, h + \sin^2\theta \, y_*^2 \, \tilde{h}\,\ \text{ and }\ h_2 \equiv h -  y_*^2 \, \tilde{h}\,,$
one of the equations decouples and we obtain
\begin{subequations}
\label{eq:eom_mass_eigenstates}
\begin{align}
h_1''+2 \, H h_1' + k^2 h_1 &= 0\,,\\
h_2''+2 \,H h_2' + k^2 h_2 + a^2 m_g^2 \, h_2 &= a^2 m_g^2 \kappa(\theta, y_*) h_1\,,
\end{align}
\end{subequations}
where we have defined the physical graviton mass $m_g^2 \nolinebreak\equiv \nolinebreak m^2 \Gamma_* (\sin^2\theta + \frac{\cos^2\theta}{y_*^2})$, and the source term is proportional to $\kappa(\theta, y_*) \equiv ( 1-y_*^2)/(\cos^2\theta + y_*^2 \sin^2\theta)$. We observe that Eqs.~\eqref{eq:eom_mass_eigenstates} comprise one massless and one massive propagating tensor perturbation, where the latter is sourced by the former. Ignoring the Hubble rate, which is typically much smaller than the wave numbers $k$ under consideration, and setting the expansion rate to a constant, $a=1$, we can solve these equations and rotate back to the physical basis,
\begin{subequations}
\begin{align}
  h(t, k) =& \frac{\cos^2\theta\cos\left(k\, t\right)+ y_*^2 \sin^2\theta \cos\left(\sqrt{k^2+m_g^2}\,t\right)}{\cos^2\theta + y_*^2 \sin^2\theta},\label{eq:physicalSolution}\\ 
  \tilde{h}(t,k) =& \frac{\cos^2\theta\cos\left(k\, t\right) - \cos^2\theta \cos\left(\sqrt{k^2+m_g^2}\,t\right)}{\cos^2\theta + y_*^2 \sin^2\theta},
\end{align}
\end{subequations}
where $\eta$ has been replaced by cosmic time $t$ as per $a=1$.

\begin{figure*}[tb]
\centering
\subfloat[$m_g = 10^{-22}\eV$]{\includegraphics[width=.49\textwidth]{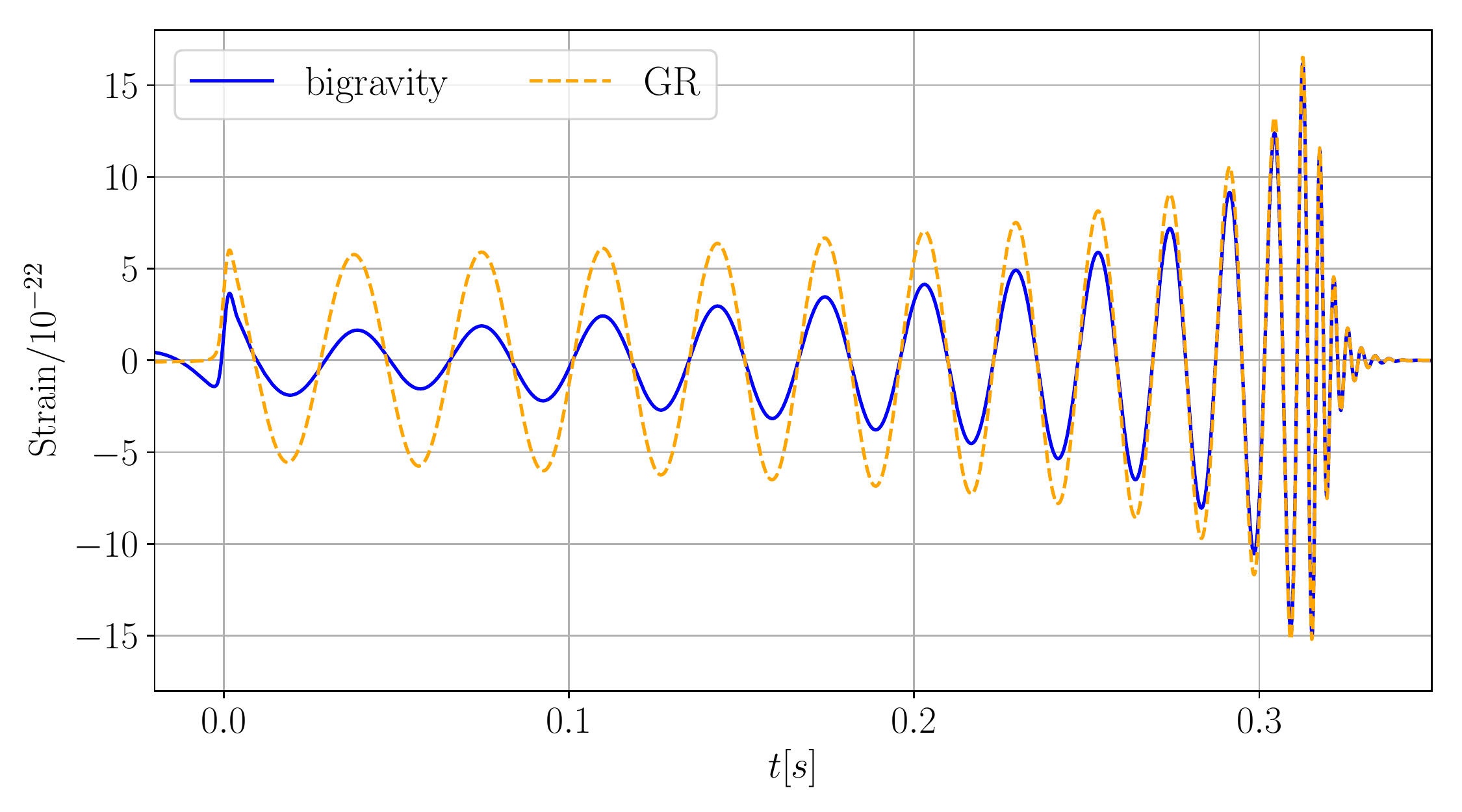}}
\hfill
\subfloat[$m_g = 10^{-19}\eV$]{\includegraphics[width=.49\textwidth]{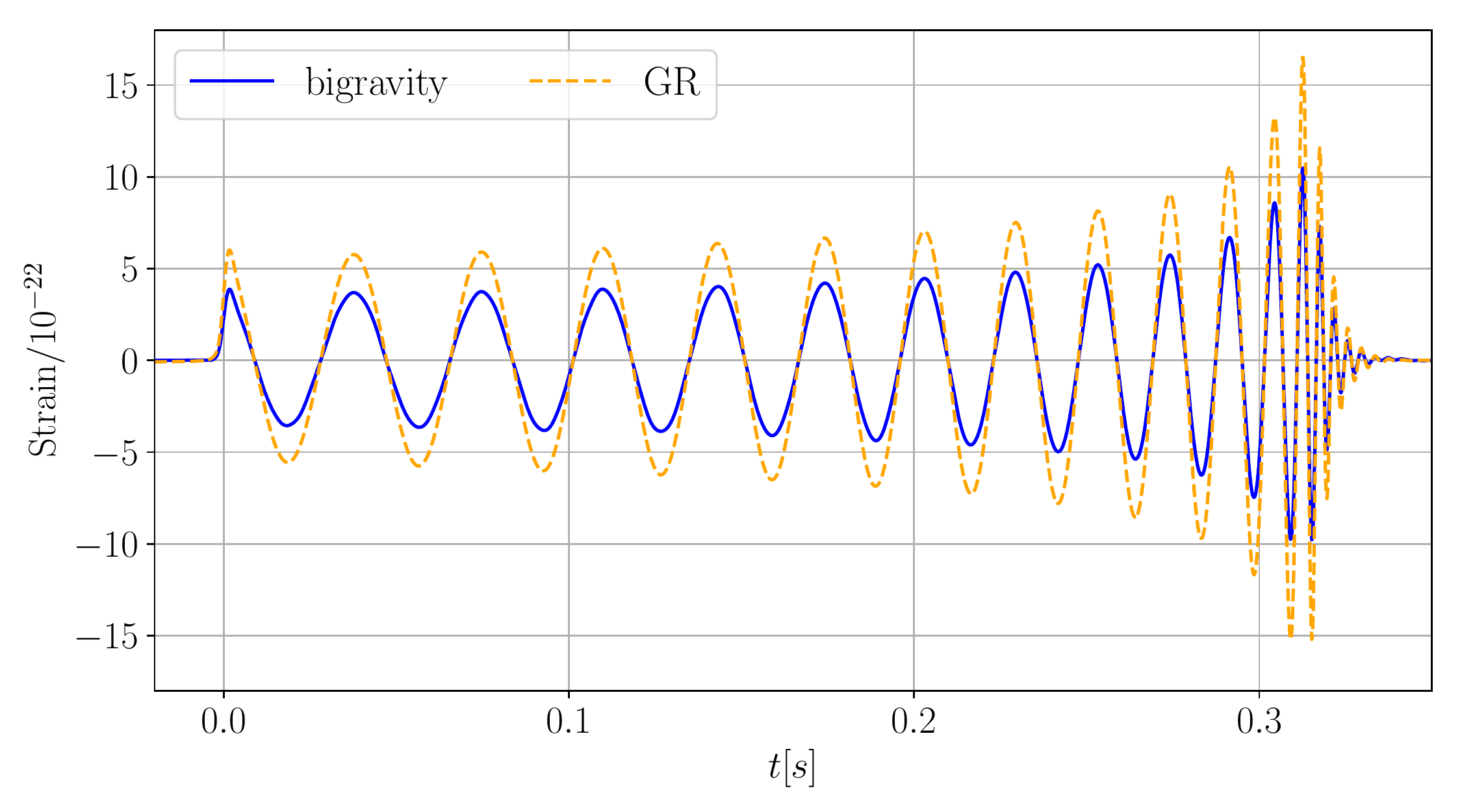}}
\caption{\label{fig:StrainModulation}Bigravity vs.\ GR: simulated strain in the detector due to gravitational waves as emitted by the black hole merger event GW150914. The dashed orange curve shows the results in GR, while the solid blue curve is obtained by multiplying with the frequency-dependent modulation due to bigravity. Note the constant suppression in panel (b).}
\end{figure*}

Since the graviton mass is restricted to be much smaller than the typical wave number $k$, we may expand $\sqrt{k^2+m_g^2} \simeq k \left[ 1 + {m_g^2}/ {(2 k^2)} \right] \equiv \omega_0 + \delta \omega$. We see that the numerator in Eq.~\eqref{eq:physicalSolution} is minimized when the second cosine acquires a total phase shift of $\delta \omega\, T_*\, \pi$, and thus,
$  T_* (\omega_0) = \frac{2\pi\, \omega_0}{m_g^2}\,,$
which coincides with the expression for the oscillation length for neutrinos, confirming our na\"ive expectation.

In order to make a quantitative statement about the modulation of the strain observed in GW observations, we average this expression over a timescale $T$, which is bigger than the period of one massless mode's inverse frequency, $T_0 = \frac{2\pi}{\omega_0}$, but much smaller than the period of the modulation induced by the mass term, $T_* = \frac{\pi}{\delta \omega}$. Squaring the strain, we find its envelope function
where the normalization is determined by the condition $\left.\left\langle h^2(t,k)\right\rangle\right|_{T=0} =1$, i.e.~initially a pure perturbation of the physical metric has been excited.

Finally, we aim to express the strain in terms of the cosmic redshift $z$, which is defined as $1+z = a(t_0)/a(t)$. For a universe dominated by a CC, we find that $H = \mathrm{const.}$ and $a(t) = e^{H t}$. We therefore express the time as $t = -\frac{1}{H} \log (1+z)$.\footnote{Note that we have reinstated $a(t)\neq$ const.~in conflict with the condition $a=1$ used in the analytic derivation of Eq.~\eqref{eq:solutionz}. Thus Eq.~\eqref{eq:solutionz} is only a valid approximation for small $z$.} In summary, the squared amplitude of the GW signal in bigravity is modulated as
\begin{widetext}
\begin{equation}
\label{eq:solutionz}
  \left\langle h^2(z,k)\right\rangle_{T_0 \ll T \ll T_*} =  \frac{\cos^4\theta}{ \left(\cos^2\theta + y_*^2 \sin^2\theta \right)^2} \left[ 1 + y_*^4 \tan^4\theta + 2 \,y_*^2 \tan^2\theta \,\cos \left(\frac{\delta \omega}{H}\, \log(1+z)\right) \right].
\end{equation}
\end{widetext}
At this point, we would like to point out that the phenomenon has previously been studied in~\cite{DeFelice2014,Narikawa:2014fua}, where the authors find a modulation that is proportional to $\tilde{c}-1$. As we will outline in the following, this is not the leading effect in our analysis, where oscillations occur also in flat space. Furthermore, we find that the phenomenon leads to a reduction rather than amplification of the amplitude compared to GR, as expected from neutrino oscillations. Both are physically sensible outcomes.

\subsection{Phenomenology} \label{sec:pheno}

\begin{figure}[!hb]
\centering
\includegraphics[width=.49\textwidth]{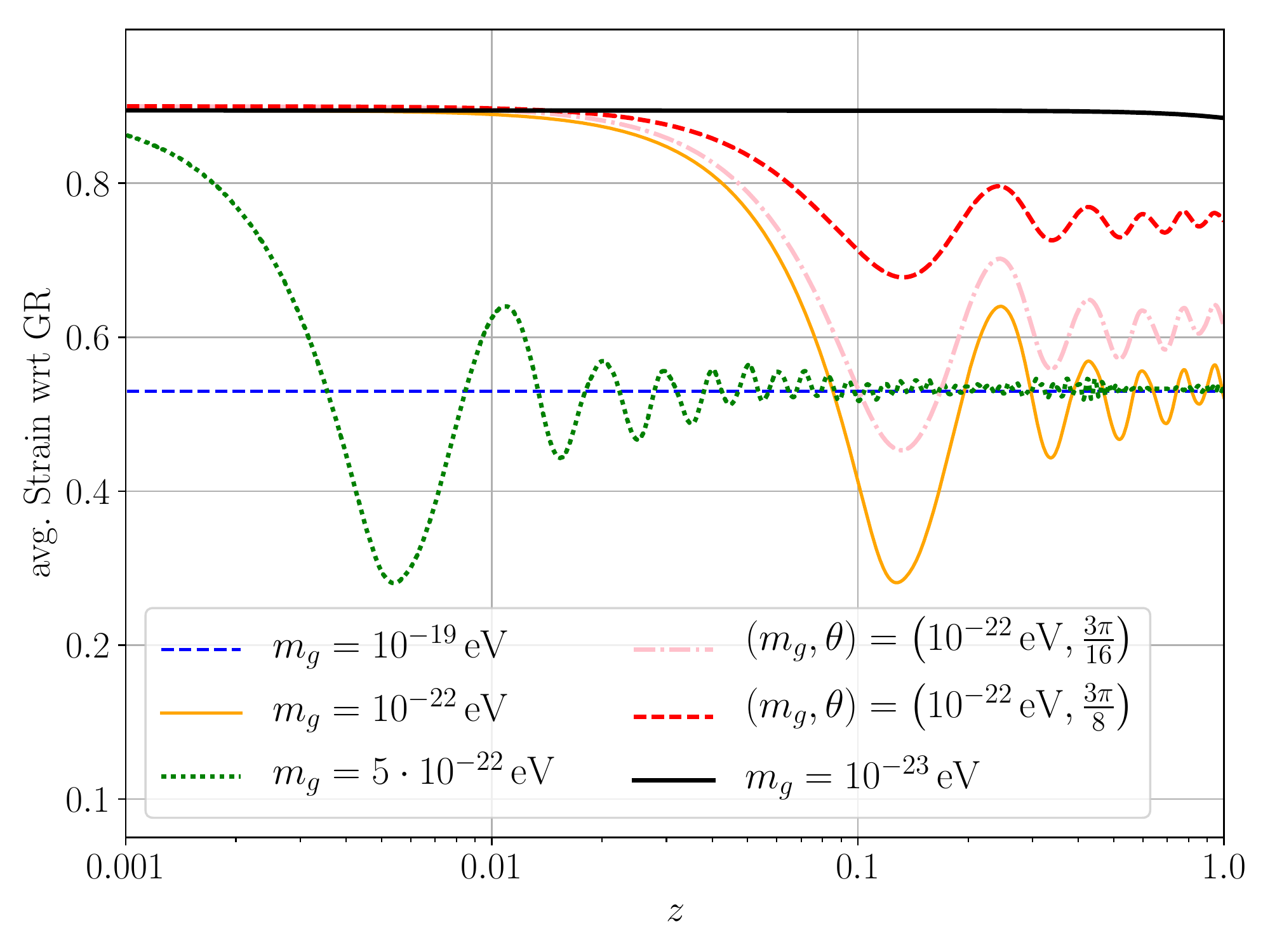}
\caption{\label{fig:zPlots}Average suppression of a GW150914-like strain as a function of the redshift for different sets of the parameters $m_g$ and $\theta$ ($= \pi/4$ unless stated explicitly). Note that, for large $m_g$ and redshift, the suppression levels out at $\sim 64\%$ as discussed in the main text for $\theta = \pi/4$. The value of the mixing angle $\theta$ determines the average level of reduction of the strain relative to GR at large distances.}
\end{figure}

Given that we have reached a quantitative understanding of GW oscillations in terms of the modulation~\eqref{eq:solutionz}, we now ask whether this effect is visible in realistic scenarios. To this end, we have made use of the available data for the events GW150914~\cite{Abbott:2016izl} and GW151226~\cite{Abbott:2016nmj} obtained by means of numerical simulations~\cite{EinsteinToolkit1,EinsteinToolkit2,EinsteinToolkit3,EinsteinToolkit4,EinsteinToolkit5,EinsteinToolkit6,EinsteinToolkit7,EinsteinToolkit8,EinsteinToolkit9,EinsteinToolkit10,EinsteinToolkit11,SXS:catalog}. This yields the strain as it would be observed in a detector on Earth. We then modulate the strain according to Eq.~\eqref{eq:solutionz}. Two such examples for GW150914 are shown in Fig.~\ref{fig:StrainModulation}, where the parameters are chosen such that one obtains a maximally visible effect, i.e.~$\theta=\pi/4$ and $y_*=1$. One observes that a graviton mass of $m_g= 10^{-22}\eV$ strongly changes the shape of the signal, where the modulation is at first strongly suppressing the amplitude and then gradually approaching the GR amplitude towards the typical merger peak, commonly referred to as \emph{chirp}. On the other hand, a larger graviton mass $m_g=10^{-19}\eV$ leads to a global suppression of the amplitude by a constant factor. This effect is similar to the decoherence of oscillating neutrino wave packets and we will now briefly discuss this effect.

The massive and the massless modes propagate in wave packets with different group velocities $v_g = \frac{\partial \omega}{\partial k}$. As for very light, relativistic neutrinos, the difference of group velocities is approximately given by $\Delta v_g \simeq \frac{m_g^2}{2 E^2}$. The wave packets will de-cohere, i.e.~interference will be absent and the frequency dependence of the suppression is lost to a constant reduction, once the time of propagation exceeds $T_\text{coh} \sim L_\text{coh} / c \sim \sigma_x / \Delta v_g$,
where $\sigma_x$ is the spatial/temporal width of the wave packet \cite{Beuthe:2001rc}. Since its determination would involve an exact solution of the full set of Einstein equations for the system, it will be practically impossible to obtain $\sigma_x$. However, from the shape of the signal, we estimate $\sigma_x \sim 0.1\, \mathrm{s}$ for GW150914. Therefore, we find that for $E / \hbar \sim 100 \, \mathrm{Hz}$,
\begin{equation}\label{eq:decoherence_Length}
	L_\text{coh} \sim 0.1\, \mathrm{s}\ \frac{ 2 E^2}{m_g^2} = \left( \frac{10^{-22}\, \mathrm{eV}}{m_g}\right)^2  \mathrm{Gpc}\,.
\end{equation}
This rather heuristic argument is nevertheless in good agreement with Fig.~\ref{fig:zPlots}, where for $m_g = 10^{-22}\eV$ no averaging is observable at distances of the order 100 Mpc, while for $m_g = 5 \cdot 10^{-22}\eV$, or even $m_g = 10^{-19} \eV$, the amplitude levels out for distances below the Gpc scale. Note that the longer time scale of GW151226 has little effect on the mass scale relevant for decoherence by virtue of Eq.~\eqref{eq:decoherence_Length}. The resulting $\mathcal{O}(1)$ correction is not relevant for the estimate presented here.

Once the distance increases beyond the scale set by $L_\text{coh}$, the strain suppression relative to the prediction of GR caused by oscillations levels out. E.g.~for $y_*=1$, $\theta=\pi/4$ we find
$\left\langle h(t,k)\right\rangle_{T \gg T_\text{coh}} =  \frac{2}{\pi}$,
which predicts a suppression factor constant in frequency and distance of about 64 \% at large redshifts, which is clearly confirmed in Fig.~\ref{fig:zPlots}.

Note that higher graviton masses lead to shorter length scales before the amplitude averages out, in complete analogy to neutrino oscillations. In practice, such a frequency-independent suppression is indistinguishable from ordinary GWs of GR and would be interpreted as a larger redshift, i.e.~one would generally \emph{overestimate} the redshift on such BBH merger events. However if the source of the GW can be localized e.g.~by electromagnetic observations, a discrepancy between the inferred redshift and the one obtained from the GW amplitude within GR could hint at graviton oscillations in the decoherence regime. Additionally, if a larger set of events becomes available, this can be used to constrain the larger-graviton mass regime by comparing expected distribution of BBH systems with the observed event rates. 

\begin{figure}[h]
\centering
\includegraphics[width=.45\textwidth]{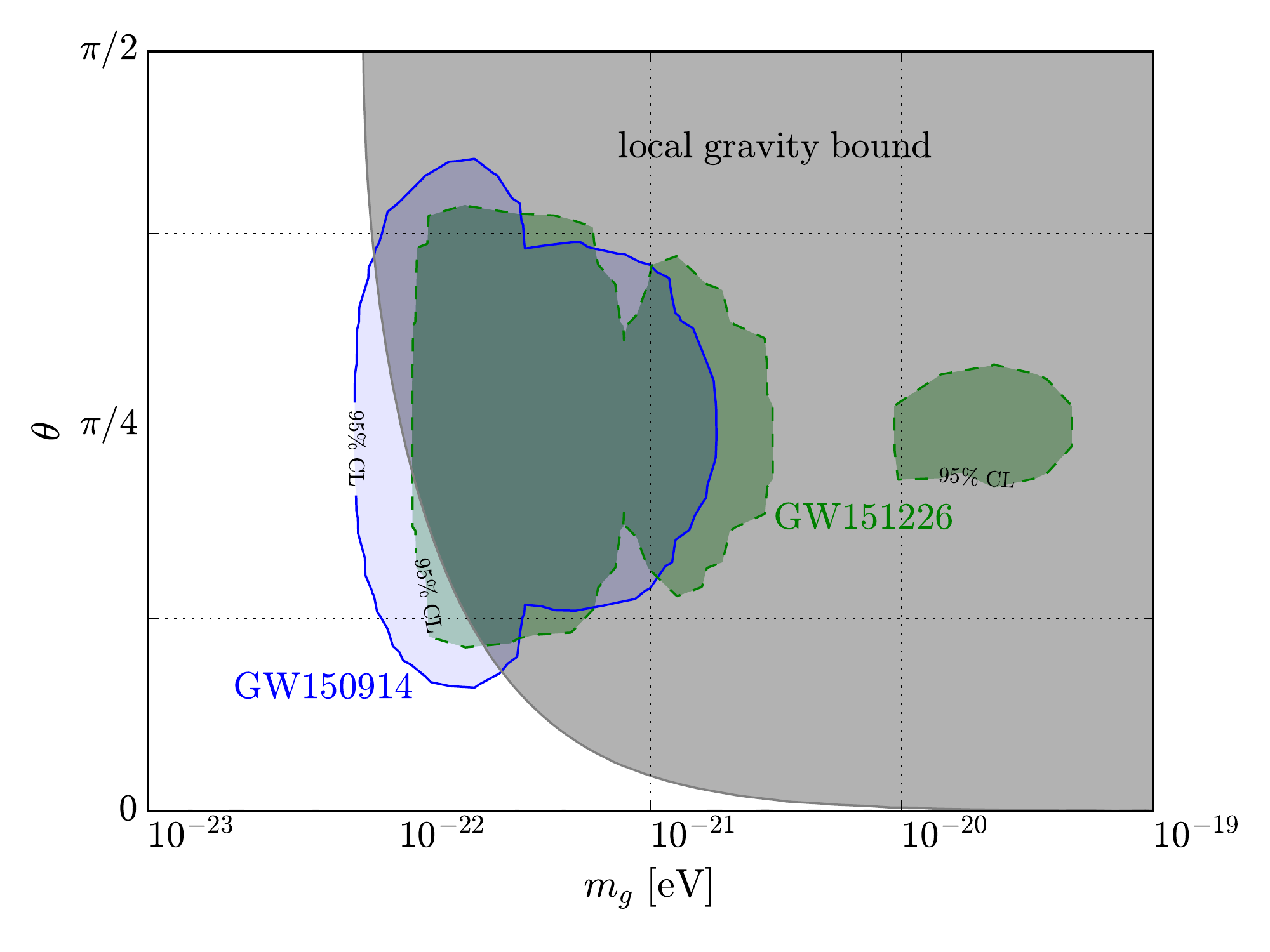}
\caption{\label{fig:LimitsGW150914}Excluded parameter space due to a simplified wave form analysis as discussed in the main text. Note that massive gravity is recovered for ${\theta  =  \pi/2}$, from which we apply model independent mass bounds.}
\end{figure}

For the low-mass regime, we can constrain the parameters of the model by demanding that the waveform be in agreement with the error bars of the observed events. We have used a simple $\chi^2$-analysis to obtain Fig.~\ref{fig:LimitsGW150914}, where we set $y_*=1$ exploiting the parameter redundancy of $m$ and the $\beta_i$.
For very small $m_g$, or $\theta\approx 0, \pi/2$, the suppression vanishes. Similarly, all events that lie beyond $L_\text{coh}$ are
indistinguishable from an equivalent event in GR at larger~$z$. 
From the remaining events the waveform in bigravity is clearly distinguishable from the GR strain, and we draw our conclusions on the excluded parameter space. We note that GW150914 gives stronger constraints than the second event GW151226. But even with only one observation, we find that for large enough mixing angles we may exclude values of $m_g \gtrsim 10^{-22}\eV$, comparable to the bounds set by GW150914 via a modified dispersion relation~\cite{Abbott:2016blz}. 
We have adopted the model-independent mass bound from solar system tests, $ m_g < 7.2 \cdot 10^{-23}\eV$~\cite{deRham:2016nuf}, to the present case by multiplying the mass with a factor $\sin \theta$ to account for the bigravity modification of the classical Newtonian potential, see e.g.~\cite{Platscher:2016adw}. We find that GW oscillations give stronger constraints for smaller mixing angles, where the bound from local gravity tests quickly becomes weaker.
In conclusion, GW oscillations offer excellent prospects to probe the bigravity parameter space once more events at higher precision become available.


\section{Conclusions} \label{sec:Conclusions}

We have studied the oscillatory behavior of gravitational waves in the framework of bigravity
In full analogy to neutrino oscillations, we have seen that a non-diagonal coupling of the two modes to matter gives rise to potentially significant modulations of the strain that would be observable e.g.~in the LIGO or LISA detectors. Using the first ever detected gravitational wave signals GW150914 and GW151226, we illustrated that the bigravity modification of GR can lead to drastic  modulations of the strain compared to the predictions of GR. Using this, we have constrained the parameter space of the model in the low-mass regime, and pointed out that, once more events are available, the high-mass regime can be constrained, too.

In this letter, we have 
made several approximations and assumptions in order to be able to give compact analytic expressions that allow the reader to understand the mechanisms behind gravitational wave oscillations. Nevertheless, the fully general results are obtained easily by following our approach such that future analyses may directly use the results of this work.

\section*{acknowledgments}
We would like to thank Evgeny Akhmedov for very useful discussions on the fundamentals of neutrino oscillations. We are also grateful to Angnis~Schmidt-May and Mikael~von~Strauss for very useful comments on the manuscript. MP is supported by IMPRS-PTFS.

\section*{Note added}
Simultaneously to this manuscript, \cite{Brax:2017hxh} appeared, where GW oscillations in doubly coupled bigravity are studied. Note that there, the leading effect is proportional to $\tilde{c}-1$ because of the democratic coupling of the tensors to matter.

\phantom{.}

\bibliography{literature}

\begin{thebibliography}{50}%
\makeatletter
\providecommand \@ifxundefined [1]{%
 \@ifx{#1\undefined}
}%
\providecommand \@ifnum [1]{%
 \ifnum #1\expandafter \@firstoftwo
 \else \expandafter \@secondoftwo
 \fi
}%
\providecommand \@ifx [1]{%
 \ifx #1\expandafter \@firstoftwo
 \else \expandafter \@secondoftwo
 \fi
}%
\providecommand \natexlab [1]{#1}%
\providecommand \enquote  [1]{``#1''}%
\providecommand \bibnamefont  [1]{#1}%
\providecommand \bibfnamefont [1]{#1}%
\providecommand \citenamefont [1]{#1}%
\providecommand \href@noop [0]{\@secondoftwo}%
\providecommand \href [0]{\begingroup \@sanitize@url \@href}%
\providecommand \@href[1]{\@@startlink{#1}\@@href}%
\providecommand \@@href[1]{\endgroup#1\@@endlink}%
\providecommand \@sanitize@url [0]{\catcode `\\12\catcode `\$12\catcode
  `\&12\catcode `\#12\catcode `\^12\catcode `\_12\catcode `\%12\relax}%
\providecommand \@@startlink[1]{}%
\providecommand \@@endlink[0]{}%
\providecommand \url  [0]{\begingroup\@sanitize@url \@url }%
\providecommand \@url [1]{\endgroup\@href {#1}{\urlprefix }}%
\providecommand \urlprefix  [0]{URL }%
\providecommand \Eprint [0]{\href }%
\providecommand \doibase [0]{http://dx.doi.org/}%
\providecommand \selectlanguage [0]{\@gobble}%
\providecommand \bibinfo  [0]{\@secondoftwo}%
\providecommand \bibfield  [0]{\@secondoftwo}%
\providecommand \translation [1]{[#1]}%
\providecommand \BibitemOpen [0]{}%
\providecommand \bibitemStop [0]{}%
\providecommand \bibitemNoStop [0]{.\EOS\space}%
\providecommand \EOS [0]{\spacefactor3000\relax}%
\providecommand \BibitemShut  [1]{\csname bibitem#1\endcsname}%
\let\auto@bib@innerbib\@empty
\bibitem [{\citenamefont {Fierz}\ and\ \citenamefont
  {Pauli}(1939)}]{Fierz:1939ix}%
  \BibitemOpen
  \bibfield  {author} {\bibinfo {author} {\bibfnamefont {M.}~\bibnamefont
  {Fierz}}\ and\ \bibinfo {author} {\bibfnamefont {W.}~\bibnamefont {Pauli}},\
  }\href {\doibase 10.1098/rspa.1939.0140} {\bibfield  {journal} {\bibinfo
  {journal} {Proc. Roy. Soc. Lond.}\ }\textbf {\bibinfo {volume} {A173}},\
  \bibinfo {pages} {211} (\bibinfo {year} {1939})}\BibitemShut {NoStop}%
\bibitem [{\citenamefont {Fierz}(1939)}]{Fierz:1939zz}%
  \BibitemOpen
  \bibfield  {author} {\bibinfo {author} {\bibfnamefont {M.}~\bibnamefont
  {Fierz}},\ }\href@noop {} {\bibfield  {journal} {\bibinfo  {journal} {Helv.
  Phys. Acta}\ }\textbf {\bibinfo {volume} {12}},\ \bibinfo {pages} {3}
  (\bibinfo {year} {1939})}\BibitemShut {NoStop}%
\bibitem [{\citenamefont {van Dam}\ and\ \citenamefont
  {Veltman}(1970)}]{vanDam:1970vg}%
  \BibitemOpen
  \bibfield  {author} {\bibinfo {author} {\bibfnamefont {H.}~\bibnamefont {van
  Dam}}\ and\ \bibinfo {author} {\bibfnamefont {M.~J.~G.}\ \bibnamefont
  {Veltman}},\ }\href {\doibase 10.1016/0550-3213(70)90416-5} {\bibfield
  {journal} {\bibinfo  {journal} {Nucl. Phys.}\ }\textbf {\bibinfo {volume}
  {B22}},\ \bibinfo {pages} {397} (\bibinfo {year} {1970})}\BibitemShut
  {NoStop}%
\bibitem [{\citenamefont {Zakharov}(1970)}]{Zakharov:1970cc}%
  \BibitemOpen
  \bibfield  {author} {\bibinfo {author} {\bibfnamefont {V.~I.}\ \bibnamefont
  {Zakharov}},\ }\href@noop {} {\bibfield  {journal} {\bibinfo  {journal} {JETP
  Lett.}\ }\textbf {\bibinfo {volume} {12}},\ \bibinfo {pages} {312} (\bibinfo
  {year} {1970})},\ \bibinfo {note} {[Pisma Zh. Eksp. Teor.
  Fiz.12,447(1970)]}\BibitemShut {NoStop}%
\bibitem [{\citenamefont {Vainshtein}(1972)}]{Vainshtein:1972sx}%
  \BibitemOpen
  \bibfield  {author} {\bibinfo {author} {\bibfnamefont {A.~I.}\ \bibnamefont
  {Vainshtein}},\ }\href {\doibase 10.1016/0370-2693(72)90147-5} {\bibfield
  {journal} {\bibinfo  {journal} {Phys. Lett.}\ }\textbf {\bibinfo {volume}
  {B39}},\ \bibinfo {pages} {393} (\bibinfo {year} {1972})}\BibitemShut
  {NoStop}%
\bibitem [{\citenamefont {Boulware}\ and\ \citenamefont
  {Deser}(1972)}]{Boulware:1973my}%
  \BibitemOpen
  \bibfield  {author} {\bibinfo {author} {\bibfnamefont {D.~G.}\ \bibnamefont
  {Boulware}}\ and\ \bibinfo {author} {\bibfnamefont {S.}~\bibnamefont
  {Deser}},\ }\href {\doibase 10.1103/PhysRevD.6.3368} {\bibfield  {journal}
  {\bibinfo  {journal} {Phys. Rev.}\ }\textbf {\bibinfo {volume} {D6}},\
  \bibinfo {pages} {3368} (\bibinfo {year} {1972})}\BibitemShut {NoStop}%
\bibitem [{\citenamefont {de~Rham}\ and\ \citenamefont
  {Gabadadze}(2010)}]{deRham:2010ik}%
  \BibitemOpen
  \bibfield  {author} {\bibinfo {author} {\bibfnamefont {C.}~\bibnamefont
  {de~Rham}}\ and\ \bibinfo {author} {\bibfnamefont {G.}~\bibnamefont
  {Gabadadze}},\ }\href {\doibase 10.1103/PhysRevD.82.044020} {\bibfield
  {journal} {\bibinfo  {journal} {Phys. Rev.}\ }\textbf {\bibinfo {volume}
  {D82}},\ \bibinfo {pages} {044020} (\bibinfo {year} {2010})},\ \Eprint
  {http://arxiv.org/abs/1007.0443} {arXiv:1007.0443 [hep-th]} \BibitemShut
  {NoStop}%
\bibitem [{\citenamefont {de~Rham}\ \emph {et~al.}(2011)\citenamefont
  {de~Rham}, \citenamefont {Gabadadze},\ and\ \citenamefont
  {Tolley}}]{deRham:2010kj}%
  \BibitemOpen
  \bibfield  {author} {\bibinfo {author} {\bibfnamefont {C.}~\bibnamefont
  {de~Rham}}, \bibinfo {author} {\bibfnamefont {G.}~\bibnamefont {Gabadadze}},
  \ and\ \bibinfo {author} {\bibfnamefont {A.~J.}\ \bibnamefont {Tolley}},\
  }\href {\doibase 10.1103/PhysRevLett.106.231101} {\bibfield  {journal}
  {\bibinfo  {journal} {Phys. Rev. Lett.}\ }\textbf {\bibinfo {volume} {106}},\
  \bibinfo {pages} {231101} (\bibinfo {year} {2011})},\ \Eprint
  {http://arxiv.org/abs/1011.1232} {arXiv:1011.1232 [hep-th]} \BibitemShut
  {NoStop}%
\bibitem [{\citenamefont {de~Rham}\ \emph {et~al.}(2012)\citenamefont
  {de~Rham}, \citenamefont {Gabadadze},\ and\ \citenamefont
  {Tolley}}]{deRham:2011rn}%
  \BibitemOpen
  \bibfield  {author} {\bibinfo {author} {\bibfnamefont {C.}~\bibnamefont
  {de~Rham}}, \bibinfo {author} {\bibfnamefont {G.}~\bibnamefont {Gabadadze}},
  \ and\ \bibinfo {author} {\bibfnamefont {A.~J.}\ \bibnamefont {Tolley}},\
  }\href {\doibase 10.1016/j.physletb.2012.03.081} {\bibfield  {journal}
  {\bibinfo  {journal} {Phys. Lett.}\ }\textbf {\bibinfo {volume} {B711}},\
  \bibinfo {pages} {190} (\bibinfo {year} {2012})},\ \Eprint
  {http://arxiv.org/abs/1107.3820} {arXiv:1107.3820 [hep-th]} \BibitemShut
  {NoStop}%
\bibitem [{\citenamefont {Hassan}\ and\ \citenamefont
  {Rosen}(2012{\natexlab{a}})}]{Hassan:2011hr}%
  \BibitemOpen
  \bibfield  {author} {\bibinfo {author} {\bibfnamefont {S.~F.}\ \bibnamefont
  {Hassan}}\ and\ \bibinfo {author} {\bibfnamefont {R.~A.}\ \bibnamefont
  {Rosen}},\ }\href {\doibase 10.1103/PhysRevLett.108.041101} {\bibfield
  {journal} {\bibinfo  {journal} {Phys. Rev. Lett.}\ }\textbf {\bibinfo
  {volume} {108}},\ \bibinfo {pages} {041101} (\bibinfo {year}
  {2012}{\natexlab{a}})},\ \Eprint {http://arxiv.org/abs/1106.3344}
  {arXiv:1106.3344 [hep-th]} \BibitemShut {NoStop}%
\bibitem [{\citenamefont {Hassan}\ and\ \citenamefont
  {Rosen}(2011)}]{Hassan:2011vm}%
  \BibitemOpen
  \bibfield  {author} {\bibinfo {author} {\bibfnamefont {S.~F.}\ \bibnamefont
  {Hassan}}\ and\ \bibinfo {author} {\bibfnamefont {R.~A.}\ \bibnamefont
  {Rosen}},\ }\href {\doibase 10.1007/JHEP07(2011)009} {\bibfield  {journal}
  {\bibinfo  {journal} {JHEP}\ }\textbf {\bibinfo {volume} {07}},\ \bibinfo
  {pages} {009} (\bibinfo {year} {2011})},\ \Eprint
  {http://arxiv.org/abs/1103.6055} {arXiv:1103.6055 [hep-th]} \BibitemShut
  {NoStop}%
\bibitem [{\citenamefont {Hassan}\ \emph {et~al.}(2012)\citenamefont {Hassan},
  \citenamefont {Rosen},\ and\ \citenamefont {Schmidt-May}}]{Hassan:2011tf}%
  \BibitemOpen
  \bibfield  {author} {\bibinfo {author} {\bibfnamefont {S.~F.}\ \bibnamefont
  {Hassan}}, \bibinfo {author} {\bibfnamefont {R.~A.}\ \bibnamefont {Rosen}}, \
  and\ \bibinfo {author} {\bibfnamefont {A.}~\bibnamefont {Schmidt-May}},\
  }\href {\doibase 10.1007/JHEP02(2012)026} {\bibfield  {journal} {\bibinfo
  {journal} {JHEP}\ }\textbf {\bibinfo {volume} {02}},\ \bibinfo {pages} {026}
  (\bibinfo {year} {2012})},\ \Eprint {http://arxiv.org/abs/1109.3230}
  {arXiv:1109.3230 [hep-th]} \BibitemShut {NoStop}%
\bibitem [{\citenamefont {Comelli}\ \emph
  {et~al.}(2012{\natexlab{a}})\citenamefont {Comelli}, \citenamefont
  {Crisostomi}, \citenamefont {Nesti},\ and\ \citenamefont
  {Pilo}}]{Comelli:2012vz}%
  \BibitemOpen
  \bibfield  {author} {\bibinfo {author} {\bibfnamefont {D.}~\bibnamefont
  {Comelli}}, \bibinfo {author} {\bibfnamefont {M.}~\bibnamefont {Crisostomi}},
  \bibinfo {author} {\bibfnamefont {F.}~\bibnamefont {Nesti}}, \ and\ \bibinfo
  {author} {\bibfnamefont {L.}~\bibnamefont {Pilo}},\ }\href {\doibase
  10.1103/PhysRevD.86.101502} {\bibfield  {journal} {\bibinfo  {journal} {Phys.
  Rev.}\ }\textbf {\bibinfo {volume} {D86}},\ \bibinfo {pages} {101502}
  (\bibinfo {year} {2012}{\natexlab{a}})},\ \Eprint
  {http://arxiv.org/abs/1204.1027} {arXiv:1204.1027 [hep-th]} \BibitemShut
  {NoStop}%
\bibitem [{\citenamefont {Deffayet}\ \emph
  {et~al.}(2013{\natexlab{a}})\citenamefont {Deffayet}, \citenamefont
  {Mourad},\ and\ \citenamefont {Zahariade}}]{Deffayet:2012nr}%
  \BibitemOpen
  \bibfield  {author} {\bibinfo {author} {\bibfnamefont {C.}~\bibnamefont
  {Deffayet}}, \bibinfo {author} {\bibfnamefont {J.}~\bibnamefont {Mourad}}, \
  and\ \bibinfo {author} {\bibfnamefont {G.}~\bibnamefont {Zahariade}},\ }\href
  {\doibase 10.1088/1475-7516/2013/01/032} {\bibfield  {journal} {\bibinfo
  {journal} {JCAP}\ }\textbf {\bibinfo {volume} {1301}},\ \bibinfo {pages}
  {032} (\bibinfo {year} {2013}{\natexlab{a}})},\ \Eprint
  {http://arxiv.org/abs/1207.6338} {arXiv:1207.6338 [hep-th]} \BibitemShut
  {NoStop}%
\bibitem [{\citenamefont {Deffayet}\ \emph
  {et~al.}(2013{\natexlab{b}})\citenamefont {Deffayet}, \citenamefont
  {Mourad},\ and\ \citenamefont {Zahariade}}]{Deffayet:2012zc}%
  \BibitemOpen
  \bibfield  {author} {\bibinfo {author} {\bibfnamefont {C.}~\bibnamefont
  {Deffayet}}, \bibinfo {author} {\bibfnamefont {J.}~\bibnamefont {Mourad}}, \
  and\ \bibinfo {author} {\bibfnamefont {G.}~\bibnamefont {Zahariade}},\ }\href
  {\doibase 10.1007/JHEP03(2013)086} {\bibfield  {journal} {\bibinfo  {journal}
  {JHEP}\ }\textbf {\bibinfo {volume} {03}},\ \bibinfo {pages} {086} (\bibinfo
  {year} {2013}{\natexlab{b}})},\ \Eprint {http://arxiv.org/abs/1208.4493}
  {arXiv:1208.4493 [gr-qc]} \BibitemShut {NoStop}%
\bibitem [{\citenamefont {de~Rham}\ \emph {et~al.}(2016)\citenamefont
  {de~Rham}, \citenamefont {Tolley},\ and\ \citenamefont
  {Zhou}}]{deRham:2016plk}%
  \BibitemOpen
  \bibfield  {author} {\bibinfo {author} {\bibfnamefont {C.}~\bibnamefont
  {de~Rham}}, \bibinfo {author} {\bibfnamefont {A.~J.}\ \bibnamefont {Tolley}},
  \ and\ \bibinfo {author} {\bibfnamefont {S.-Y.}\ \bibnamefont {Zhou}},\
  }\href {\doibase 10.1007/JHEP04(2016)188} {\bibfield  {journal} {\bibinfo
  {journal} {JHEP}\ }\textbf {\bibinfo {volume} {04}},\ \bibinfo {pages} {188}
  (\bibinfo {year} {2016})},\ \Eprint {http://arxiv.org/abs/1602.03721}
  {arXiv:1602.03721 [hep-th]} \BibitemShut {NoStop}%
\bibitem [{\citenamefont {Hassan}\ and\ \citenamefont
  {Rosen}(2012{\natexlab{b}})}]{Hassan:2011zd}%
  \BibitemOpen
  \bibfield  {author} {\bibinfo {author} {\bibfnamefont {S.~F.}\ \bibnamefont
  {Hassan}}\ and\ \bibinfo {author} {\bibfnamefont {R.~A.}\ \bibnamefont
  {Rosen}},\ }\href {\doibase 10.1007/JHEP02(2012)126} {\bibfield  {journal}
  {\bibinfo  {journal} {JHEP}\ }\textbf {\bibinfo {volume} {02}},\ \bibinfo
  {pages} {126} (\bibinfo {year} {2012}{\natexlab{b}})},\ \Eprint
  {http://arxiv.org/abs/1109.3515} {arXiv:1109.3515 [hep-th]} \BibitemShut
  {NoStop}%
\bibitem [{\citenamefont {Hassan}\ and\ \citenamefont
  {Rosen}(2012{\natexlab{c}})}]{Hassan:2011ea}%
  \BibitemOpen
  \bibfield  {author} {\bibinfo {author} {\bibfnamefont {S.~F.}\ \bibnamefont
  {Hassan}}\ and\ \bibinfo {author} {\bibfnamefont {R.~A.}\ \bibnamefont
  {Rosen}},\ }\href {\doibase 10.1007/JHEP04(2012)123} {\bibfield  {journal}
  {\bibinfo  {journal} {JHEP}\ }\textbf {\bibinfo {volume} {04}},\ \bibinfo
  {pages} {123} (\bibinfo {year} {2012}{\natexlab{c}})},\ \Eprint
  {http://arxiv.org/abs/1111.2070} {arXiv:1111.2070 [hep-th]} \BibitemShut
  {NoStop}%
\bibitem [{\citenamefont {de~Rham}\ \emph {et~al.}(2014)\citenamefont
  {de~Rham}, \citenamefont {Heisenberg},\ and\ \citenamefont
  {Ribeiro}}]{deRham:2014fha}%
  \BibitemOpen
  \bibfield  {author} {\bibinfo {author} {\bibfnamefont {C.}~\bibnamefont
  {de~Rham}}, \bibinfo {author} {\bibfnamefont {L.}~\bibnamefont {Heisenberg}},
  \ and\ \bibinfo {author} {\bibfnamefont {R.~H.}\ \bibnamefont {Ribeiro}},\
  }\href {\doibase 10.1103/PhysRevD.90.124042} {\bibfield  {journal} {\bibinfo
  {journal} {Phys. Rev.}\ }\textbf {\bibinfo {volume} {D90}},\ \bibinfo {pages}
  {124042} (\bibinfo {year} {2014})},\ \Eprint {http://arxiv.org/abs/1409.3834}
  {arXiv:1409.3834 [hep-th]} \BibitemShut {NoStop}%
\bibitem [{\citenamefont {Berezhiani}\ \emph {et~al.}(2007)\citenamefont
  {Berezhiani}, \citenamefont {Comelli}, \citenamefont {Nesti},\ and\
  \citenamefont {Pilo}}]{Berezhiani:2007zf}%
  \BibitemOpen
  \bibfield  {author} {\bibinfo {author} {\bibfnamefont {Z.}~\bibnamefont
  {Berezhiani}}, \bibinfo {author} {\bibfnamefont {D.}~\bibnamefont {Comelli}},
  \bibinfo {author} {\bibfnamefont {F.}~\bibnamefont {Nesti}}, \ and\ \bibinfo
  {author} {\bibfnamefont {L.}~\bibnamefont {Pilo}},\ }\href {\doibase
  10.1103/PhysRevLett.99.131101} {\bibfield  {journal} {\bibinfo  {journal}
  {Phys. Rev. Lett.}\ }\textbf {\bibinfo {volume} {99}},\ \bibinfo {pages}
  {131101} (\bibinfo {year} {2007})},\ \Eprint
  {http://arxiv.org/abs/hep-th/0703264} {arXiv:hep-th/0703264 [HEP-TH]}
  \BibitemShut {NoStop}%
\bibitem [{\citenamefont {Hassan}\ \emph {et~al.}(2013)\citenamefont {Hassan},
  \citenamefont {Schmidt-May},\ and\ \citenamefont {von
  Strauss}}]{Hassan:2012wr}%
  \BibitemOpen
  \bibfield  {author} {\bibinfo {author} {\bibfnamefont {S.~F.}\ \bibnamefont
  {Hassan}}, \bibinfo {author} {\bibfnamefont {A.}~\bibnamefont {Schmidt-May}},
  \ and\ \bibinfo {author} {\bibfnamefont {M.}~\bibnamefont {von Strauss}},\
  }\href {\doibase 10.1007/JHEP05(2013)086} {\bibfield  {journal} {\bibinfo
  {journal} {JHEP}\ }\textbf {\bibinfo {volume} {05}},\ \bibinfo {pages} {086}
  (\bibinfo {year} {2013})},\ \Eprint {http://arxiv.org/abs/1208.1515}
  {arXiv:1208.1515 [hep-th]} \BibitemShut {NoStop}%
\bibitem [{\citenamefont {De~Felice}\ \emph
  {et~al.}(2014{\natexlab{a}})\citenamefont {De~Felice}, \citenamefont
  {Nakamura},\ and\ \citenamefont {Tanaka}}]{DeFelice2014}%
  \BibitemOpen
  \bibfield  {author} {\bibinfo {author} {\bibfnamefont {A.}~\bibnamefont
  {De~Felice}}, \bibinfo {author} {\bibfnamefont {T.}~\bibnamefont {Nakamura}},
  \ and\ \bibinfo {author} {\bibfnamefont {T.}~\bibnamefont {Tanaka}},\ }\href
  {\doibase 10.1093/ptep/ptu024} {\bibfield  {journal} {\bibinfo  {journal}
  {Progress of Theoretical and Experimental Physics}\ }\textbf {\bibinfo
  {volume} {2014}},\ \bibinfo {pages} {1} (\bibinfo {year}
  {2014}{\natexlab{a}})}\BibitemShut {NoStop}%
\bibitem [{\citenamefont {Narikawa}\ \emph {et~al.}(2015)\citenamefont
  {Narikawa}, \citenamefont {Ueno}, \citenamefont {Tagoshi}, \citenamefont
  {Tanaka}, \citenamefont {Kanda},\ and\ \citenamefont
  {Nakamura}}]{Narikawa:2014fua}%
  \BibitemOpen
  \bibfield  {author} {\bibinfo {author} {\bibfnamefont {T.}~\bibnamefont
  {Narikawa}}, \bibinfo {author} {\bibfnamefont {K.}~\bibnamefont {Ueno}},
  \bibinfo {author} {\bibfnamefont {H.}~\bibnamefont {Tagoshi}}, \bibinfo
  {author} {\bibfnamefont {T.}~\bibnamefont {Tanaka}}, \bibinfo {author}
  {\bibfnamefont {N.}~\bibnamefont {Kanda}}, \ and\ \bibinfo {author}
  {\bibfnamefont {T.}~\bibnamefont {Nakamura}},\ }\href {\doibase
  10.1103/PhysRevD.91.062007} {\bibfield  {journal} {\bibinfo  {journal} {Phys.
  Rev.}\ }\textbf {\bibinfo {volume} {D91}},\ \bibinfo {pages} {062007}
  (\bibinfo {year} {2015})},\ \Eprint {http://arxiv.org/abs/1412.8074}
  {arXiv:1412.8074 [gr-qc]} \BibitemShut {NoStop}%
\bibitem [{\citenamefont {Abbott}\ \emph
  {et~al.}(2016{\natexlab{a}})\citenamefont {Abbott} \emph
  {et~al.}}]{Abbott:2016blz}%
  \BibitemOpen
  \bibfield  {author} {\bibinfo {author} {\bibfnamefont {B.~P.}\ \bibnamefont
  {Abbott}} \emph {et~al.} (\bibinfo {collaboration} {Virgo, LIGO
  Scientific}),\ }\href {\doibase 10.1103/PhysRevLett.116.061102} {\bibfield
  {journal} {\bibinfo  {journal} {Phys. Rev. Lett.}\ }\textbf {\bibinfo
  {volume} {116}},\ \bibinfo {pages} {061102} (\bibinfo {year}
  {2016}{\natexlab{a}})},\ \Eprint {http://arxiv.org/abs/1602.03837}
  {arXiv:1602.03837 [gr-qc]} \BibitemShut {NoStop}%
\bibitem [{\citenamefont {De~Felice}\ \emph
  {et~al.}(2014{\natexlab{b}})\citenamefont {De~Felice}, \citenamefont
  {Gümrükçüoğlu}, \citenamefont {Mukohyama}, \citenamefont {Tanahashi},\
  and\ \citenamefont {Tanaka}}]{DeFelice:2014nja}%
  \BibitemOpen
  \bibfield  {author} {\bibinfo {author} {\bibfnamefont {A.}~\bibnamefont
  {De~Felice}}, \bibinfo {author} {\bibfnamefont {A.~E.}\ \bibnamefont
  {Gümrükçüoğlu}}, \bibinfo {author} {\bibfnamefont {S.}~\bibnamefont
  {Mukohyama}}, \bibinfo {author} {\bibfnamefont {N.}~\bibnamefont
  {Tanahashi}}, \ and\ \bibinfo {author} {\bibfnamefont {T.}~\bibnamefont
  {Tanaka}},\ }\href {\doibase 10.1088/1475-7516/2014/06/037} {\bibfield
  {journal} {\bibinfo  {journal} {JCAP}\ }\textbf {\bibinfo {volume} {1406}},\
  \bibinfo {pages} {037} (\bibinfo {year} {2014}{\natexlab{b}})},\ \Eprint
  {http://arxiv.org/abs/1404.0008} {arXiv:1404.0008 [hep-th]} \BibitemShut
  {NoStop}%
\bibitem [{\citenamefont {Akrami}\ \emph {et~al.}(2015)\citenamefont {Akrami},
  \citenamefont {Hassan}, \citenamefont {Könnig}, \citenamefont
  {Schmidt-May},\ and\ \citenamefont {Solomon}}]{Akrami:2015qga}%
  \BibitemOpen
  \bibfield  {author} {\bibinfo {author} {\bibfnamefont {Y.}~\bibnamefont
  {Akrami}}, \bibinfo {author} {\bibfnamefont {S.~F.}\ \bibnamefont {Hassan}},
  \bibinfo {author} {\bibfnamefont {F.}~\bibnamefont {Könnig}}, \bibinfo
  {author} {\bibfnamefont {A.}~\bibnamefont {Schmidt-May}}, \ and\ \bibinfo
  {author} {\bibfnamefont {A.~R.}\ \bibnamefont {Solomon}},\ }\href {\doibase
  10.1016/j.physletb.2015.06.062} {\bibfield  {journal} {\bibinfo  {journal}
  {Phys. Lett.}\ }\textbf {\bibinfo {volume} {B748}},\ \bibinfo {pages} {37}
  (\bibinfo {year} {2015})},\ \Eprint {http://arxiv.org/abs/1503.07521}
  {arXiv:1503.07521 [gr-qc]} \BibitemShut {NoStop}%
\bibitem [{\citenamefont {Babichev}\ \emph {et~al.}(2016)\citenamefont
  {Babichev}, \citenamefont {Marzola}, \citenamefont {Raidal}, \citenamefont
  {Schmidt-May}, \citenamefont {Urban}, \citenamefont {{Veermäe}},\ and\
  \citenamefont {von Strauss}}]{Babichev:2016bxi}%
  \BibitemOpen
  \bibfield  {author} {\bibinfo {author} {\bibfnamefont {E.}~\bibnamefont
  {Babichev}}, \bibinfo {author} {\bibfnamefont {L.}~\bibnamefont {Marzola}},
  \bibinfo {author} {\bibfnamefont {M.}~\bibnamefont {Raidal}}, \bibinfo
  {author} {\bibfnamefont {A.}~\bibnamefont {Schmidt-May}}, \bibinfo {author}
  {\bibfnamefont {F.}~\bibnamefont {Urban}}, \bibinfo {author} {\bibfnamefont
  {H.}~\bibnamefont {{Veermäe}}}, \ and\ \bibinfo {author} {\bibfnamefont
  {M.}~\bibnamefont {von Strauss}},\ }\href {\doibase
  10.1088/1475-7516/2016/09/016} {\bibfield  {journal} {\bibinfo  {journal}
  {JCAP}\ }\textbf {\bibinfo {volume} {1609}},\ \bibinfo {pages} {016}
  (\bibinfo {year} {2016})},\ \Eprint {http://arxiv.org/abs/1607.03497}
  {arXiv:1607.03497 [hep-th]} \BibitemShut {NoStop}%
\bibitem [{\citenamefont {von Strauss}\ \emph {et~al.}(2012)\citenamefont {von
  Strauss}, \citenamefont {Schmidt-May}, \citenamefont {Enander}, \citenamefont
  {Mortsell},\ and\ \citenamefont {Hassan}}]{vonStrauss:2011mq}%
  \BibitemOpen
  \bibfield  {author} {\bibinfo {author} {\bibfnamefont {M.}~\bibnamefont {von
  Strauss}}, \bibinfo {author} {\bibfnamefont {A.}~\bibnamefont {Schmidt-May}},
  \bibinfo {author} {\bibfnamefont {J.}~\bibnamefont {Enander}}, \bibinfo
  {author} {\bibfnamefont {E.}~\bibnamefont {Mortsell}}, \ and\ \bibinfo
  {author} {\bibfnamefont {S.~F.}\ \bibnamefont {Hassan}},\ }\href {\doibase
  10.1088/1475-7516/2012/03/042} {\bibfield  {journal} {\bibinfo  {journal}
  {JCAP}\ }\textbf {\bibinfo {volume} {1203}},\ \bibinfo {pages} {042}
  (\bibinfo {year} {2012})},\ \Eprint {http://arxiv.org/abs/1111.1655}
  {arXiv:1111.1655 [gr-qc]} \BibitemShut {NoStop}%
\bibitem [{\citenamefont {Comelli}\ \emph
  {et~al.}(2012{\natexlab{b}})\citenamefont {Comelli}, \citenamefont
  {Crisostomi}, \citenamefont {Nesti},\ and\ \citenamefont
  {Pilo}}]{Comelli:2011zm}%
  \BibitemOpen
  \bibfield  {author} {\bibinfo {author} {\bibfnamefont {D.}~\bibnamefont
  {Comelli}}, \bibinfo {author} {\bibfnamefont {M.}~\bibnamefont {Crisostomi}},
  \bibinfo {author} {\bibfnamefont {F.}~\bibnamefont {Nesti}}, \ and\ \bibinfo
  {author} {\bibfnamefont {L.}~\bibnamefont {Pilo}},\ }\href {\doibase
  10.1007/JHEP06(2012)020, 10.1007/JHEP03(2012)067} {\bibfield  {journal}
  {\bibinfo  {journal} {JHEP}\ }\textbf {\bibinfo {volume} {03}},\ \bibinfo
  {pages} {067} (\bibinfo {year} {2012}{\natexlab{b}})},\ \bibinfo {note}
  {[Erratum: JHEP06,020(2012)]},\ \Eprint {http://arxiv.org/abs/1111.1983}
  {arXiv:1111.1983 [hep-th]} \BibitemShut {NoStop}%
\bibitem [{\citenamefont {Babichev}\ \emph {et~al.}(2008)\citenamefont
  {Babichev}, \citenamefont {Mukhanov},\ and\ \citenamefont
  {Vikman}}]{Babichev:2007dw}%
  \BibitemOpen
  \bibfield  {author} {\bibinfo {author} {\bibfnamefont {E.}~\bibnamefont
  {Babichev}}, \bibinfo {author} {\bibfnamefont {V.}~\bibnamefont {Mukhanov}},
  \ and\ \bibinfo {author} {\bibfnamefont {A.}~\bibnamefont {Vikman}},\ }\href
  {\doibase 10.1088/1126-6708/2008/02/101} {\bibfield  {journal} {\bibinfo
  {journal} {JHEP}\ }\textbf {\bibinfo {volume} {02}},\ \bibinfo {pages} {101}
  (\bibinfo {year} {2008})},\ \Eprint {http://arxiv.org/abs/0708.0561}
  {arXiv:0708.0561 [hep-th]} \BibitemShut {NoStop}%
\bibitem [{\citenamefont {de~Rham}(2014)}]{deRham:2014zqa}%
  \BibitemOpen
  \bibfield  {author} {\bibinfo {author} {\bibfnamefont {C.}~\bibnamefont
  {de~Rham}},\ }\href {\doibase 10.12942/lrr-2014-7} {\bibfield  {journal}
  {\bibinfo  {journal} {Living Rev. Rel.}\ }\textbf {\bibinfo {volume} {17}},\
  \bibinfo {pages} {7} (\bibinfo {year} {2014})},\ \Eprint
  {http://arxiv.org/abs/1401.4173} {arXiv:1401.4173 [hep-th]} \BibitemShut
  {NoStop}%
\bibitem [{\citenamefont {Comelli}\ \emph
  {et~al.}(2012{\natexlab{c}})\citenamefont {Comelli}, \citenamefont
  {Crisostomi},\ and\ \citenamefont {Pilo}}]{Comelli:2012db}%
  \BibitemOpen
  \bibfield  {author} {\bibinfo {author} {\bibfnamefont {D.}~\bibnamefont
  {Comelli}}, \bibinfo {author} {\bibfnamefont {M.}~\bibnamefont {Crisostomi}},
  \ and\ \bibinfo {author} {\bibfnamefont {L.}~\bibnamefont {Pilo}},\ }\href
  {\doibase 10.1007/JHEP06(2012)085} {\bibfield  {journal} {\bibinfo  {journal}
  {JHEP}\ }\textbf {\bibinfo {volume} {06}},\ \bibinfo {pages} {085} (\bibinfo
  {year} {2012}{\natexlab{c}})},\ \Eprint {http://arxiv.org/abs/1202.1986}
  {arXiv:1202.1986 [hep-th]} \BibitemShut {NoStop}%
\bibitem [{\citenamefont {Abbott}\ \emph
  {et~al.}(2016{\natexlab{b}})\citenamefont {Abbott} \emph
  {et~al.}}]{Abbott:2016izl}%
  \BibitemOpen
  \bibfield  {author} {\bibinfo {author} {\bibfnamefont {B.}~\bibnamefont
  {Abbott}} \emph {et~al.} (\bibinfo {collaboration} {Virgo, LIGO
  Scientific}),\ }\href {\doibase 10.1103/PhysRevX.6.041014} {\bibfield
  {journal} {\bibinfo  {journal} {Phys. Rev.}\ }\textbf {\bibinfo {volume}
  {X6}},\ \bibinfo {pages} {041014} (\bibinfo {year} {2016}{\natexlab{b}})},\
  \Eprint {http://arxiv.org/abs/1606.01210} {arXiv:1606.01210 [gr-qc]}
  \BibitemShut {NoStop}%
\bibitem [{\citenamefont {Abbott}\ \emph
  {et~al.}(2016{\natexlab{c}})\citenamefont {Abbott} \emph
  {et~al.}}]{Abbott:2016nmj}%
  \BibitemOpen
  \bibfield  {author} {\bibinfo {author} {\bibfnamefont {B.~P.}\ \bibnamefont
  {Abbott}} \emph {et~al.} (\bibinfo {collaboration} {Virgo, LIGO
  Scientific}),\ }\href {\doibase 10.1103/PhysRevLett.116.241103} {\bibfield
  {journal} {\bibinfo  {journal} {Phys. Rev. Lett.}\ }\textbf {\bibinfo
  {volume} {116}},\ \bibinfo {pages} {241103} (\bibinfo {year}
  {2016}{\natexlab{c}})},\ \Eprint {http://arxiv.org/abs/1606.04855}
  {arXiv:1606.04855 [gr-qc]} \BibitemShut {NoStop}%
\bibitem [{\citenamefont {Wardell}\ \emph {et~al.}(2016)\citenamefont
  {Wardell}, \citenamefont {Hinder},\ and\ \citenamefont
  {Bentivegna}}]{EinsteinToolkit1}%
  \BibitemOpen
  \bibfield  {author} {\bibinfo {author} {\bibfnamefont {B.}~\bibnamefont
  {Wardell}}, \bibinfo {author} {\bibfnamefont {I.}~\bibnamefont {Hinder}}, \
  and\ \bibinfo {author} {\bibfnamefont {E.}~\bibnamefont {Bentivegna}},\
  }\href {\doibase 10.5281/zenodo.155394} {\enquote {\bibinfo {title}
  {{Simulation of GW150914 binary black hole merger using the Einstein
  Toolkit}},}\ } (\bibinfo {year} {2016})\BibitemShut {NoStop}%
\bibitem [{\citenamefont {Loffler}\ \emph {et~al.}(2012)\citenamefont {Loffler}
  \emph {et~al.}}]{EinsteinToolkit2}%
  \BibitemOpen
  \bibfield  {author} {\bibinfo {author} {\bibfnamefont {F.}~\bibnamefont
  {Loffler}} \emph {et~al.},\ }\href {\doibase 10.1088/0264-9381/29/11/115001}
  {\bibfield  {journal} {\bibinfo  {journal} {Class. Quant. Grav.}\ }\textbf
  {\bibinfo {volume} {29}},\ \bibinfo {pages} {115001} (\bibinfo {year}
  {2012})},\ \Eprint {http://arxiv.org/abs/1111.3344} {arXiv:1111.3344 [gr-qc]}
  \BibitemShut {NoStop}%
\bibitem [{\citenamefont {Pollney}\ \emph {et~al.}(2011)\citenamefont
  {Pollney}, \citenamefont {Reisswig}, \citenamefont {Schnetter}, \citenamefont
  {Dorband},\ and\ \citenamefont {Diener}}]{EinsteinToolkit3}%
  \BibitemOpen
  \bibfield  {author} {\bibinfo {author} {\bibfnamefont {D.}~\bibnamefont
  {Pollney}}, \bibinfo {author} {\bibfnamefont {C.}~\bibnamefont {Reisswig}},
  \bibinfo {author} {\bibfnamefont {E.}~\bibnamefont {Schnetter}}, \bibinfo
  {author} {\bibfnamefont {N.}~\bibnamefont {Dorband}}, \ and\ \bibinfo
  {author} {\bibfnamefont {P.}~\bibnamefont {Diener}},\ }\href {\doibase
  10.1103/PhysRevD.83.044045} {\bibfield  {journal} {\bibinfo  {journal} {Phys.
  Rev.}\ }\textbf {\bibinfo {volume} {D83}},\ \bibinfo {pages} {044045}
  (\bibinfo {year} {2011})},\ \Eprint {http://arxiv.org/abs/0910.3803}
  {arXiv:0910.3803 [gr-qc]} \BibitemShut {NoStop}%
\bibitem [{\citenamefont {Schnetter}\ \emph {et~al.}(2004)\citenamefont
  {Schnetter}, \citenamefont {Hawley},\ and\ \citenamefont
  {Hawke}}]{EinsteinToolkit4}%
  \BibitemOpen
  \bibfield  {author} {\bibinfo {author} {\bibfnamefont {E.}~\bibnamefont
  {Schnetter}}, \bibinfo {author} {\bibfnamefont {S.~H.}\ \bibnamefont
  {Hawley}}, \ and\ \bibinfo {author} {\bibfnamefont {I.}~\bibnamefont
  {Hawke}},\ }\href {\doibase 10.1088/0264-9381/21/6/014} {\bibfield  {journal}
  {\bibinfo  {journal} {Class. Quant. Grav.}\ }\textbf {\bibinfo {volume}
  {21}},\ \bibinfo {pages} {1465} (\bibinfo {year} {2004})},\ \Eprint
  {http://arxiv.org/abs/gr-qc/0310042} {arXiv:gr-qc/0310042 [gr-qc]}
  \BibitemShut {NoStop}%
\bibitem [{\citenamefont {Thornburg}(2004)}]{EinsteinToolkit5}%
  \BibitemOpen
  \bibfield  {author} {\bibinfo {author} {\bibfnamefont {J.}~\bibnamefont
  {Thornburg}},\ }\href {\doibase 10.1088/0264-9381/21/2/026} {\bibfield
  {journal} {\bibinfo  {journal} {Class. Quant. Grav.}\ }\textbf {\bibinfo
  {volume} {21}},\ \bibinfo {pages} {743} (\bibinfo {year} {2004})},\ \Eprint
  {http://arxiv.org/abs/gr-qc/0306056} {arXiv:gr-qc/0306056 [gr-qc]}
  \BibitemShut {NoStop}%
\bibitem [{\citenamefont {Ansorg}\ \emph {et~al.}(2004)\citenamefont {Ansorg},
  \citenamefont {Bruegmann},\ and\ \citenamefont {Tichy}}]{EinsteinToolkit6}%
  \BibitemOpen
  \bibfield  {author} {\bibinfo {author} {\bibfnamefont {M.}~\bibnamefont
  {Ansorg}}, \bibinfo {author} {\bibfnamefont {B.}~\bibnamefont {Bruegmann}}, \
  and\ \bibinfo {author} {\bibfnamefont {W.}~\bibnamefont {Tichy}},\ }\href
  {\doibase 10.1103/PhysRevD.70.064011} {\bibfield  {journal} {\bibinfo
  {journal} {Phys. Rev.}\ }\textbf {\bibinfo {volume} {D70}},\ \bibinfo {pages}
  {064011} (\bibinfo {year} {2004})},\ \Eprint
  {http://arxiv.org/abs/gr-qc/0404056} {arXiv:gr-qc/0404056 [gr-qc]}
  \BibitemShut {NoStop}%
\bibitem [{\citenamefont {Dreyer}\ \emph {et~al.}(2003)\citenamefont {Dreyer},
  \citenamefont {Krishnan}, \citenamefont {Shoemaker},\ and\ \citenamefont
  {Schnetter}}]{EinsteinToolkit7}%
  \BibitemOpen
  \bibfield  {author} {\bibinfo {author} {\bibfnamefont {O.}~\bibnamefont
  {Dreyer}}, \bibinfo {author} {\bibfnamefont {B.}~\bibnamefont {Krishnan}},
  \bibinfo {author} {\bibfnamefont {D.}~\bibnamefont {Shoemaker}}, \ and\
  \bibinfo {author} {\bibfnamefont {E.}~\bibnamefont {Schnetter}},\ }\href
  {\doibase 10.1103/PhysRevD.67.024018} {\bibfield  {journal} {\bibinfo
  {journal} {Phys. Rev.}\ }\textbf {\bibinfo {volume} {D67}},\ \bibinfo {pages}
  {024018} (\bibinfo {year} {2003})},\ \Eprint
  {http://arxiv.org/abs/gr-qc/0206008} {arXiv:gr-qc/0206008 [gr-qc]}
  \BibitemShut {NoStop}%
\bibitem [{\citenamefont {Goodale}\ \emph {et~al.}(2003)\citenamefont
  {Goodale}, \citenamefont {Allen}, \citenamefont {Lanfermann}, \citenamefont
  {Mass{\'o}}, \citenamefont {Radke}, \citenamefont {Seidel},\ and\
  \citenamefont {Shalf}}]{EinsteinToolkit8}%
  \BibitemOpen
  \bibfield  {author} {\bibinfo {author} {\bibfnamefont {T.}~\bibnamefont
  {Goodale}}, \bibinfo {author} {\bibfnamefont {G.}~\bibnamefont {Allen}},
  \bibinfo {author} {\bibfnamefont {G.}~\bibnamefont {Lanfermann}}, \bibinfo
  {author} {\bibfnamefont {J.}~\bibnamefont {Mass{\'o}}}, \bibinfo {author}
  {\bibfnamefont {T.}~\bibnamefont {Radke}}, \bibinfo {author} {\bibfnamefont
  {E.}~\bibnamefont {Seidel}}, \ and\ \bibinfo {author} {\bibfnamefont
  {J.}~\bibnamefont {Shalf}},\ }in\ \href {http://edoc.mpg.de/3341} {\emph
  {\bibinfo {booktitle} {Vector and Parallel Processing -- VECPAR'2002, 5th
  International Conference, Lecture Notes in Computer Science}}}\ (\bibinfo
  {publisher} {Springer},\ \bibinfo {address} {Berlin},\ \bibinfo {year}
  {2003})\BibitemShut {NoStop}%
\bibitem [{\citenamefont {Brown}\ \emph {et~al.}(2009)\citenamefont {Brown},
  \citenamefont {Diener}, \citenamefont {Sarbach}, \citenamefont {Schnetter},\
  and\ \citenamefont {Tiglio}}]{EinsteinToolkit9}%
  \BibitemOpen
  \bibfield  {author} {\bibinfo {author} {\bibfnamefont {J.~D.}\ \bibnamefont
  {Brown}}, \bibinfo {author} {\bibfnamefont {P.}~\bibnamefont {Diener}},
  \bibinfo {author} {\bibfnamefont {O.}~\bibnamefont {Sarbach}}, \bibinfo
  {author} {\bibfnamefont {E.}~\bibnamefont {Schnetter}}, \ and\ \bibinfo
  {author} {\bibfnamefont {M.}~\bibnamefont {Tiglio}},\ }\href {\doibase
  10.1103/PhysRevD.79.044023} {\bibfield  {journal} {\bibinfo  {journal} {Phys.
  Rev.}\ }\textbf {\bibinfo {volume} {D79}},\ \bibinfo {pages} {044023}
  (\bibinfo {year} {2009})},\ \Eprint {http://arxiv.org/abs/0809.3533}
  {arXiv:0809.3533 [gr-qc]} \BibitemShut {NoStop}%
\bibitem [{\citenamefont {Husa}\ \emph {et~al.}(2006)\citenamefont {Husa},
  \citenamefont {Hinder},\ and\ \citenamefont {Lechner}}]{EinsteinToolkit10}%
  \BibitemOpen
  \bibfield  {author} {\bibinfo {author} {\bibfnamefont {S.}~\bibnamefont
  {Husa}}, \bibinfo {author} {\bibfnamefont {I.}~\bibnamefont {Hinder}}, \ and\
  \bibinfo {author} {\bibfnamefont {C.}~\bibnamefont {Lechner}},\ }\href@noop
  {} {\bibfield  {journal} {\bibinfo  {journal} {Comput. Phys. Commun.}\
  }\textbf {\bibinfo {volume} {174}},\ \bibinfo {pages} {983} (\bibinfo {year}
  {2006})},\ \Eprint {http://arxiv.org/abs/arXiv:gr-qc/0404023}
  {arXiv:gr-qc/0404023} \BibitemShut {NoStop}%
\bibitem [{\citenamefont {Thomas}\ and\ \citenamefont
  {Schnetter}(2010)}]{EinsteinToolkit11}%
  \BibitemOpen
  \bibfield  {author} {\bibinfo {author} {\bibfnamefont {M.}~\bibnamefont
  {Thomas}}\ and\ \bibinfo {author} {\bibfnamefont {E.}~\bibnamefont
  {Schnetter}},\ }in\ \href {\doibase 10.1109/GRID.2010.5698010} {\emph
  {\bibinfo {booktitle} {Grid Computing (GRID), 2010 11th IEEE/ACM
  International Conference on}}}\ (\bibinfo {year} {2010})\ pp.\ \bibinfo
  {pages} {369 --378},\ \Eprint {http://arxiv.org/abs/arXiv:1008.4571 [cs.DC]}
  {arXiv:1008.4571 [cs.DC]} \BibitemShut {NoStop}%
\bibitem [{SXS(2016)}]{SXS:catalog}%
  \BibitemOpen
  \href@noop {} {}\bibinfo {howpublished}
  {\url{http://www.black-holes.org/waveforms}, catalog entry SXS:BBH:0317 for
  GW151226} (\bibinfo {year} {2016})\BibitemShut {NoStop}%
\bibitem [{\citenamefont {Beuthe}(2003)}]{Beuthe:2001rc}%
  \BibitemOpen
  \bibfield  {author} {\bibinfo {author} {\bibfnamefont {M.}~\bibnamefont
  {Beuthe}},\ }\href {\doibase 10.1016/S0370-1573(02)00538-0} {\bibfield
  {journal} {\bibinfo  {journal} {Phys. Rept.}\ }\textbf {\bibinfo {volume}
  {375}},\ \bibinfo {pages} {105} (\bibinfo {year} {2003})},\ \Eprint
  {http://arxiv.org/abs/hep-ph/0109119} {arXiv:hep-ph/0109119 [hep-ph]}
  \BibitemShut {NoStop}%
\bibitem [{\citenamefont {de~Rham}\ \emph {et~al.}(2017)\citenamefont
  {de~Rham}, \citenamefont {Deskins}, \citenamefont {Tolley},\ and\
  \citenamefont {Zhou}}]{deRham:2016nuf}%
  \BibitemOpen
  \bibfield  {author} {\bibinfo {author} {\bibfnamefont {C.}~\bibnamefont
  {de~Rham}}, \bibinfo {author} {\bibfnamefont {J.~T.}\ \bibnamefont
  {Deskins}}, \bibinfo {author} {\bibfnamefont {A.~J.}\ \bibnamefont {Tolley}},
  \ and\ \bibinfo {author} {\bibfnamefont {S.-Y.}\ \bibnamefont {Zhou}},\
  }\href {\doibase 10.1103/RevModPhys.89.025004} {\bibfield  {journal}
  {\bibinfo  {journal} {Rev. Mod. Phys.}\ }\textbf {\bibinfo {volume} {89}},\
  \bibinfo {pages} {025004} (\bibinfo {year} {2017})},\ \Eprint
  {http://arxiv.org/abs/1606.08462} {arXiv:1606.08462 [astro-ph.CO]}
  \BibitemShut {NoStop}%
\bibitem [{\citenamefont {Platscher}\ and\ \citenamefont
  {Smirnov}(2017)}]{Platscher:2016adw}%
  \BibitemOpen
  \bibfield  {author} {\bibinfo {author} {\bibfnamefont {M.}~\bibnamefont
  {Platscher}}\ and\ \bibinfo {author} {\bibfnamefont {J.}~\bibnamefont
  {Smirnov}},\ }\href {\doibase 10.1088/1475-7516/2017/03/051} {\bibfield
  {journal} {\bibinfo  {journal} {JCAP}\ }\textbf {\bibinfo {volume} {1703}},\
  \bibinfo {pages} {051} (\bibinfo {year} {2017})},\ \Eprint
  {http://arxiv.org/abs/1611.09385} {arXiv:1611.09385 [gr-qc]} \BibitemShut
  {NoStop}%
\bibitem [{\citenamefont {Brax}\ \emph {et~al.}(2017)\citenamefont {Brax},
  \citenamefont {Davis},\ and\ \citenamefont {Noller}}]{Brax:2017hxh}%
  \BibitemOpen
  \bibfield  {author} {\bibinfo {author} {\bibfnamefont {P.}~\bibnamefont
  {Brax}}, \bibinfo {author} {\bibfnamefont {A.-C.}\ \bibnamefont {Davis}}, \
  and\ \bibinfo {author} {\bibfnamefont {J.}~\bibnamefont {Noller}},\
  }\href@noop {} {\  (\bibinfo {year} {2017})},\ \Eprint
  {http://arxiv.org/abs/1703.08016} {arXiv:1703.08016 [gr-qc]} \BibitemShut
  {NoStop}%
\end{thebibliography}%

\end{document}